\documentclass{appolb}
\usepackage{epsfig}
\begin{document}

% \eqsec  % uncomment this line to get equations numbered by (sec.num)

% preprint numbers at line 279 in appolb.cls

\title{Scalar mesons and tetraquarks from twisted mass lattice QCD\thanks{Presented at ``Excited QCD 2013'', Bjelasnica Mountain, Sarajevo.}}

\author{
Marc Wagner$^1$, Constantia Alexandrou$^{2,3}$, \\ Jan Oliver Daldrop$^4$, Mattia Dalla Brida$^5$, Mario Gravina$^2$, Luigi Scorzato$^6$, Carsten Urbach$^4$, Christian Wiese$^{7,8}$
\address{
$^1$~Goethe-Universit\"at Frankfurt am Main, Institut f\"ur Theoretische Physik, Max-von-Laue-Stra{\ss}e 1, D-60438 Frankfurt am Main, Germany \\
$^2$~Department of Physics, University of Cyprus, P.O.\ Box 20537, 1678 Nicosia, Cyprus \\
$^3$~Computation-based Science and Technology Research Center, Cyprus Institute, 20 Kavafi Str., Nicosia 2121, Cyprus \\
$^4$~Helmholtz-Institut f{\"u}r Strahlen- und Kernphysik (Theorie) and Bethe Center for Theoretical Physics, Universit{\"a}t Bonn, D-53115 Bonn, Germany \\
$^5$~School of Mathematics, Trinity College, Dublin 2, Ireland \\
$^6$~ECT$^\star$, Strada delle Tabarelle, 286, I-38123, Trento, Italy \\
$^7$~NIC, DESY Zeuthen, Platanenallee 6, D-15738 Zeuthen, Germany \\
$^8$~Humboldt-Universit\"at zu Berlin, Institut f\"ur Physik, Newtonstra{\ss}e 15, \\ D-12489 Berlin, Germany
}
}

\maketitle

\vspace{-0.7cm}
\begin{center}
\includegraphics{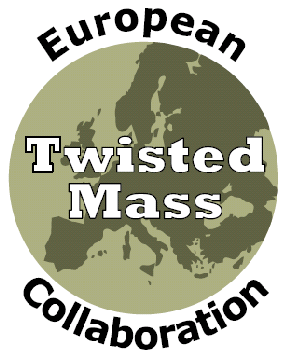}
\end{center}

\begin{abstract}
We study light scalar mesons with particular focus on the $a_0(980)$ using lattice QCD with 2+1+1 dynamical quark flavors. To investigate the structure of these scalar mesons and to identify, whether a sizeable tetraquark component is present, we use a large set of operators, including diquark-antidiquark, mesonic molecule and two-meson operators. We find that the low-lying states overlap essentially exclusively with two-meson states. This indicates that in the channels investigated no tightly bound four-quark states of either molecular or diquark-antidiquark type exist.
\end{abstract}

% 12.38.Gc 	Lattice QCD calculations (see also 11.15.Ha Lattice gauge theory)
% 14.40.Be 	Light mesons (S=C=B=0)
% 14.40.Df 	Strange mesons (|S|>0, C=B=0)


% ********************
% ********************
% ********************
% ********************
% ********************

\section{Introduction}

The nonet of light scalar mesons formed by $\sigma \equiv f_0(500)$, $\kappa \equiv K_0^\ast(800)$, $a_0(980)$ and $f_0(980)$ is poorly understood. Compared to expectation all nine states are rather light and their ordering is inverted, which might indicate a strong tetraquark component.
% For example in a standard quark antiquark picture the $a_0(980)$ states, which have $I = 1$, must necessarily be composed of two light quarks, e.g.\ $a_0(980) \equiv \bar{d} u$, while the $\kappa$ states with $I = 1/2$ are made from a strange and a light quark, e.g.\ $\kappa \equiv \bar{s} u$. Consequently, $\kappa$ should be heavier than $a_0(980)$, since $m_s > m_{u,d}$. In experiments, however, the opposite is observed, i.e.\ $m(\kappa) = 682 \pm 29 \, \textrm{MeV}$, while $m(a_0(980)) = 980 \pm 20 \, \textrm{MeV}$. On the other hand in a four-quark or tetraquark picture the quark content could be $a_0(980) \equiv \bar{d} u \bar{s} s$ and \\ $\kappa \equiv \bar{s} u (\bar{u} u + \bar{d} d)$ naturally explaining the observed ordering. Moreover, certain decay channels, e.g.\ $a_0(980) \rightarrow K + \bar{K}$, indicate that besides the two light quarks also an $s \bar{s}$ pair is present and, therefore, also support a tetraquark interpretation.
A detailed discussion of light scalar mesons can be found in \cite{Jaffe:2004ph}. They also have been discussed extensively on this conference (cf.\ e.g.\ the related publications \cite{Pelaez:2013jp,Parganlija:2012fy}). There are also various tetraquark candidates among the heavy mesons, e.g.\ the rather light $D_{s0}^\ast(2317)$ and $D_{s1}(2460)$ mesons, whose masses seem to be difficult to reproduce theoretically using standard quark antiquark computations (cf.\ e.g.\ \cite{Ebert:2009ua,Mohler:2011ke,Kalinowski:2012re}).

Here we report about the status of an ongoing long-term lattice QCD project with the aim to study possible tetraquark candidates from first principles. The focus of this talk is on the $a_0(980)$. Parts of this work have already been published \cite{Daldrop:2012sr,Alexandrou:2012rm,Wagner:2012ay}.

% ********************
% ********************
% ********************

\section{\label{SEC001}Lattice setup and four-quark creation operators}

We use gauge link configurations with 2+1+1 dynamical quark flavors generated by the European Twisted Mass Collaboration (ETMC) \cite{Baron:2008xa,Baron:2009zq,Baron:2010bv,Baron:2011sf,Baron:2010th,Baron:2010vp}. For the results shown in this talk disconnected diagrams have been ignored, which are technically rather challenging. An important physical consequence is that the quark number and the antiquark number are separately conserved for each flavor. Therefore, there is no mixing between $\bar{u} u$, $\bar{d} d$ and $\bar{s} s$ resulting in an $\eta_s$ meson with flavor structure $\bar{s} s$ instead of $\eta$ and $\eta'$ \cite{Alexandrou:2012rm}. We are currently exploring efficient techniques to compute the relevant disconnected diagrams \cite{Wagner:2012ay}.

In the following we focus on the $a_0(980)$ sector, which has quantum numbers $I(J^P) = 1(0^+)$. As usual in lattice QCD we extract the low lying spectrum in that sector by studying the asymptotic exponential behavior of Euclidean correlation functions $C_{j k}(t) \langle (\mathcal{O}_j(t))^\dagger \mathcal{O}_k(0) \rangle$. $\mathcal{O}_j$ and $\mathcal{O}_k$ denote suitable creation operators, i.e.\ operators generating the $a_0(980)$ quantum numbers, when applied to the vacuum state.

Assuming that the experimentally measured $a_0(980)$ with mass \\ $980 \pm 20 \, \textrm{MeV}$ is a rather strongly bound four quark state, suitable creation operators to excite such a state are
\begin{eqnarray}
\label{EQN001} & & \hspace{-0.7cm} \mathcal{O}_{a_0(980)}^{K \bar{K} \textrm{\scriptsize{} molecule}} \ \ = \ \ \sum_\mathbf{x} \Big(\bar{s}(\mathbf{x}) \gamma_5 u(\mathbf{x})\Big) \Big(\bar{d}(\mathbf{x}) \gamma_5 s(\mathbf{x})\Big) \\
\label{EQN002} & & \hspace{-0.7cm} \mathcal{O}_{a_0(980)}^{\textrm{\scriptsize diquark}} \ \ = \ \ \sum_\mathbf{x} \Big(\epsilon^{a b c} \bar{s}^b(\mathbf{x}) C \gamma_5 \bar{d}^{c,T}(\mathbf{x})\Big) \Big(\epsilon^{a d e} u^{d,T}(\mathbf{x}) C \gamma_5 s^e(\mathbf{x})\Big) .
\end{eqnarray}
The first operator has the spin/color structure of a $K \bar{K}$ molecule ($\bar{s}(\mathbf{x}) \gamma_5 u(\mathbf{x})$ and $\bar{d}(\mathbf{x}) \gamma_5 s(\mathbf{x})$ correspond to a kaon $K$ and an antikaon $\bar{K}$ at the same position $\mathbf{x}$). The second resembles a bound diquark antidiquark pair, where spin coupling via $C \gamma_5$ corresponds to the lightest diquarks/antidiquarks (cf.\ e.g.\ \cite{Jaffe:2004ph,Alexandrou:2006cq,Wagner:2011fs}).

Further low lying states in this sector are the two particle states $K + \bar{K}$ and $\eta_s + \pi$. Suitable creation operators to resolve these states are
\begin{eqnarray}
\label{EQN003} & & \hspace{-0.7cm} \mathcal{O}_{a_0(980)}^{K + \bar{K} \textrm{\scriptsize{} two-particle}} \ \ = \ \ \bigg(\sum_\mathbf{x} \bar{s}(\mathbf{x}) \gamma_5 u(\mathbf{x})\bigg) \bigg(\sum_\mathbf{y} \bar{d}(\mathbf{y}) \gamma_5 s(\mathbf{y})\bigg) \\
\label{EQN004} & & \hspace{-0.7cm} \mathcal{O}_{a_0(980)}^{\eta_s + \pi \textrm{\scriptsize{} two-particle}} \ \ = \ \ \bigg(\sum_\mathbf{x} \bar{s}(\mathbf{x}) \gamma_5 s(\mathbf{x})\bigg) \bigg(\sum_\mathbf{y} \bar{d}(\mathbf{y}) \gamma_5 u(\mathbf{y})\bigg) .
\end{eqnarray}

% ********************
% ********************
% ********************

\section{Numerical results an their interpretation}

We start by discussing numerical results for an ensemble with rather small spatial extension of $L \approx 1.72 \, \textrm{fm}$. This ensemble is particularly suited to distinguish two-particle states with relative momentum from states with two particles at rest and from possibly existing $a_0(980)$ tetraquark states (two-particle states with relative momentum have a rather large energy, because one quantum of momentum $p_\textrm{\scriptsize min} = 2 \pi / L \approx 720 \, \textrm{MeV}$).

\begin{figure}[t]
\includegraphics{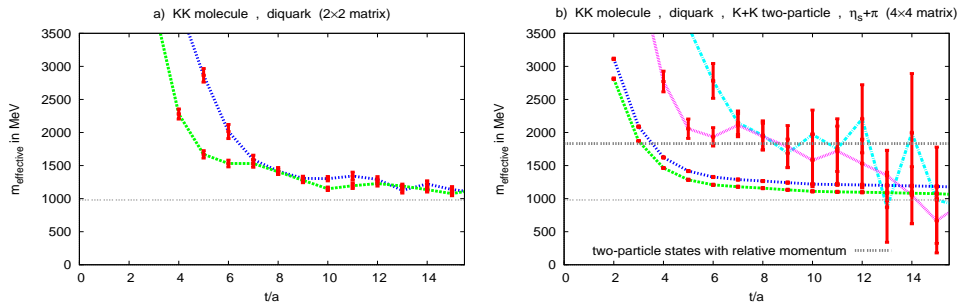}
\caption{\label{FIG001}$a_0(980)$ sector, $(L/a)^3 \times (T/a) = 20^3 \times 48$.
\textbf{a)}~Effective masses as functions of the temporal separation, $2 \times 2$ correlation matrix (operators: $K \bar{K}$ molecule, diquark-antidiquark, eqs.\ (3.2) and (3.3)).
\textbf{b)}~$4 \times 4$ correlation matrix (operators: $K \bar{K}$ molecule, diquark-antidiquark, two-particle $K + \bar{K}$, two-particle $\eta_s + \pi$, eqs.\ (3.2) to (3.5)).
%
% \textbf{c)}, \textbf{d)}~Squared eigenvector components of the two low-lying states from \textbf{b)} as functions of the temporal separation.
}
\end{figure}

Figure~\ref{FIG001}a shows effective mass plots from a $2 \times 2$ correlation matrix with a $K \bar{K}$ molecule operator (\ref{EQN001}) and a diquark-antidiquark operator (\ref{EQN002}). The corresponding two plateaus are around $1100 \, \textrm{MeV}$ and, therefore, consistent both with the expectation for possibly existing $a_0(980)$ tetra\-quark states and with two-particle $K + \bar{K}$ and $\eta_s + \pi$ states, where both particles are at rest ($m(K + \bar{K}) \approx 2 m(K) \approx 1198 \, \textrm{MeV}$; $m(\eta_s + \pi) \approx m(\eta_s) + m(\pi) \approx 1115 \, \textrm{MeV}$ in our lattice setup).

% ***
% ***
% ***

% ***
% ***
% ***

Increasing this correlation matrix to $4 \times 4$ by adding two-particle $K + \bar{K}$ and $\eta_s + \pi$ operators (eqs.\ (\ref{EQN003}) and (\ref{EQN004})) yields the effective mass results shown in Figure~\ref{FIG001}b. Two additional states are observed, whose plateaus are around $1500 \, \textrm{MeV} \ldots 2000 \, \textrm{MeV}$. From this $4 \times 4$ analysis we conclude the following:
\begin{itemize}
\item We do not observe a third low-lying state around $1100 \, \textrm{MeV}$, even though we provide operators, which are of tetraquark type as well as of two-particle type. This suggests that the two low-lying states are the expected two-particle $K + \bar{K}$ and $\eta_s + \pi$ states, while an additional stable $a_0(980)$ tetraquark state does not exist.

\item The effective masses of the two low-lying states are of much better quality in Figure~\ref{FIG001}b than in Figure~\ref{FIG001}a. We attribute this to the two-particle $K + \bar{K}$ and $\eta_s + \pi$ operators, which presumably create larger overlap to those states than the tetraquark operators. This in turn confirms the interpretation of the two observed low-lying states as two-particle states.

\item Analyzing the eigenvector components of the two low-lying states from Figure~\ref{FIG001}b we find that the lowest state is essentially exclusively of $\eta_s + \pi$ type, whereas the second lowest state is of $K + \bar{K}$ type. On the other hand, the two tetraquark operators are irrelevant for resolving those states, i.e.\ they do not seem to contribute any important structure, which is not already present in the two-particle operators. This gives additional strong support of the above interpretation of the two observed low lying states as two-particle states.

\item The energy of two-particle excitations with one relative quantum of momentum can be estimated according to \\
$m(1 + 2,p = p_\textrm{\scriptsize min}) \approx \sqrt{m(1)^2 + p_\textrm{\scriptsize min}^2} + \sqrt{m(2)^2 + p_\textrm{\scriptsize min}^2}$ with \\ $p_\textrm{\scriptsize min} = 2 \pi / L$. Inserting the meson masses corresponding to our lattice setup, $m(K) \approx 599 \, \textrm{MeV}$, $m(\eta_s) \approx 774 \, \textrm{MeV}$ and $m(\pi) \approx 341 \, \textrm{MeV}$, yields $m(K + \bar{K},p = p_\textrm{\scriptsize min}) \approx 1873 \, \textrm{MeV}$ and \\ $m(\eta_s + \pi,p = p_\textrm{\scriptsize min}) \approx 1853 \, \textrm{MeV}$. These numbers are consistent with the effective mass plateaus of the second and third excitation in Figure~\ref{FIG001}b. Consequently, we also interpret them as two-particle states.
\end{itemize}

We obtained qualitatively identical results, when varying the light quark mass and the spacetime volume \cite{Alexandrou:2012rm}.

Using exactly the same techniques, i.e.\ four-quark operators of tetraquark and of two-particle type, we also studied the $\kappa$-sector (for details cf.\ \cite{Alexandrou:2012rm}). Again we find no sign of any four-quark bound state besides the expected two-particle spectrum (in this case $K + \pi$ states). Note that this result is in contradiction to a very similar recent lattice study of the $\kappa$ meson \cite{Prelovsek:2010kg}, where an additional low lying four-quark bound state has been observed.

% ********************
% ********************
% ********************

\section{Conclusions and future plans}

We have studied the $a_0(980)$ and the $\kappa$ channel by means of 2+1+1 flavor lattice QCD using four-quark operators of molecule, diquark and two-particle type. Besides the expected two-particle spectrum (two essentially non-interacting pseudoscalar mesons) no indication of any additional low lying state, in particular no sign of a four-quark bound state could be observed. This suggests that both the $a_0(980)$ and $\kappa$ meson have either no sizeable tetraquark component or they are rather weakly bound unstable states. To investigate the latter one needs to study the volume dependence of the two-particle spectrum in the corresponding sectors (``L\"uscher's method'', cf.\ e.g.\ \cite{Luscher:1986pf,Luscher:1990ux,Luscher:1991cf}). Such computations are very challenging using lattice QCD, but first results have recently been published (cf.\ \cite{Lang:2012sv,Mohler:2012na}). We plan to perform similar computations with our setup in the near future.

% ********************
% ********************
% ********************

\section*{Acknowledgments}

M.W.\ thanks the organizers of ``Excited QCD 2013'' for the invitation to give this talk. M.W.\ acknowledges support by the Emmy Noether Programme of the DFG (German Research Foundation), grant WA 3000/1-1. M.G. acknowledges support by the Marie-Curie European training network ITN STRONGnet grant PITN-GA-2009-238353. M.D.B. is currently funded by the Irish Research Council, acknowledges support by STRONGnet and the AuroraScience project, and is grateful for the hospitality at ECT* and the University of Cyprus, where part of this work was carried out. L.S. acknowledges support from the AuroraScience project funded by the Province of Trento and INFN. This work was supported in part by the Helmholtz International Center for FAIR within the framework of the LOEWE program launched by the State of Hesse and by the DFG and the NSFC through funds provided to the Sino-German CRC 110 ``Symmetries and the Emergence of Structure in QCD''. Part of the computations presented here were performed on the Aurora system in Trento.

% ********************
% ********************
% ********************
% ********************
% ********************

% **********
% **********
% **********
% **********
% **********

\end{document}